# A portable widefield fundus camera with high dynamic range imaging capability

ALFA ROSSI,[1] MOJTABA RAHIMI,[1] DAVID LE,[1] TAEYOON SON,[1] MICHAEL J. HEIFERMAN,[2] R. V. PAUL CHAN[1,2], AND XINCHENG YAO[1,2,*]

[1]*Department of Biomedical Engineering, University of Illinois at Chicago, Chicago, IL 60607, USA*
[2]*Department of Ophthalmology and Visual Sciences, University of Illinois at Chicago, Chicago, IL 60612, USA*
*\*xcy@uic.edu*

**Abstract:** Fundus photography is indispensable for clinical detection and management of eye diseases. Limited image contrast and field of view (FOV) are common limitations of conventional fundus cameras, making it difficult to detect subtle abnormalities at the early stages of eye diseases. Further improvements of image contrast and FOV coverage are important to improve early disease detection and reliable treatment assessment. We report here a portable fundus camera, with a wide FOV and high dynamic range (HDR) imaging capabilities. Miniaturized indirect ophthalmoscopy illumination was employed to achieve the portable design for nonmydriatic, widefield fundus photography. Orthogonal polarization control was used to eliminate illumination reflectance artifact. With independent power controls, three fundus images were sequentially acquired and fused to achieve HDR function for local image contrast enhancement. A 101° eye-angle (67° visual-angle) snapshot FOV was achieved for nonmydriatic fundus photography. The effective FOV can be readily expanded up to 190° eye-angle (134° visual-angle) with the aid of a fixation target, without the need of pharmacologic pupillary dilation. The effectiveness of HDR imaging was validated with both normal healthy and pathologic eyes, compared to a conventional fundus camera.

## 1. Introduction

Vision threating eye diseases, such as diabetic retinopathy (DR), glaucoma, and age-related macular degeneration (AMD) affect approximately 400 million people worldwide, and the number of people with these conditions is projected to increase to 560 million by the year 2045 [1-3]. Early detection and prompt treatment assessment are important to prevent vision loss due to these diseases. Since eye conditions can affect both the peripheral and central retina, widefield fundus photography is important for the clinical management of eye diseases. Scanning laser ophthalmoscopy (SLO), such as Optomap (Optos, Marlborough, MA) and Eidon (Icare USA Inc., Raleigh, NC) systems can provide widefield and ultra-widefield fundus imaging [4-6] by combing two or more laser wavelengths. However, sophisticated scanning and illumination devices needed for SLOs make them typically bulky and expensive. As almost 89% of visually impaired patients are from low- and middle-income countries (LMIC) [7] with a lack of eye care providers to care for the population, a portable fundus camera to facilitate affordable telemedicine is preferable.

Currently available commercial portable fundus cameras have a limited field of view (FOV) and image contrast, which are the common limitations of annular trans-pupillary illumination [8, 9]. The FOV of conventional fundus cameras is typically around 45°-67.5° eye angle (30°-45° visual angle) [10]. In conventional fundus cameras, the illumination light is delivered through the annular shape peripherical region of the available pupil, and the imaging light from the back of the eye is collected through the center of the pupil (Fig. 1 (a)). At the pupil plane, the illumination and imaging windows must have some buffer zone in between to prevent reflection artifacts from the cornea and crystalline lens [11]. At the retina plane, the region for imaging must be covered by the illumination light delivered through the annular-shaped window. Therefore, the regions used for light illumination and imaging observation should be carefully balanced. This tradeoff limits the FOV, thus pharmacologic pupillary dilation, which is stressful for both patient and operator, is often required to expand the FOV for comprehensive eye examination [12].

Another common challenge of fundus photography is that the luminescence range needed to capture all the details in a human fundus exceeds the dynamic range of the camera sensors. For example, the light reflected from the optic nerve head is multiple order greater than that of the macular region due to the different density of pigmented cells and nerve fibers [13]. Therefore, when the illumination power is adjusted to image the peripapillary region, the macular region falls below the noise floor of the camera sensor. Adjusting the power level for better macula imaging often leads to saturation at the optic nerve head. The structural details residing below the noise floor or in the saturated region cannot be recovered. As the compromised image contrast and inhomogeneity can hinder grading of retinal disease severely [14], a more holistic image is preferable. A plethora of research has been performed to study fundus image quality improvement by correcting scene homogeneity or enhancing the contrast between retinal segments [15, 16]. These algorithms perform poorly when preserving detail in saturated regions and can often create false color that can lead to misdiagnosis.

Miniaturized indirect illumination (Fig. 1 (b)) provides a simple workaround for the limitation of the available pupil, making it an alternative solution for developing widefield portable fundus cameras [17]. In this work, we present a miniaturized indirect illumination based portable fundus camera to achieve portable, nonmydriatic, widefield fundus imaging. Orthogonal polarization control is used to eliminate the back reflected light artifacts encountered in miniaturized indirect illumination-based fundus systems [17, 18]. High dynamic range (HDR) imaging capability is integrated to enhance image contract for better visualization of fundus details with different brightness levels.

In fundus imaging, the conventional unit for FOV evaluation is the visual-angle degree. However, recently emerging widefield fundus imagers such as Optomap (Optos Inc., Marlborough, MA), ICON (Neolight, Pleasanton CA) and Retcam (Natus Medical Systems, Pleasanton, CA) is using eye-angle degree as the unit of measurement. For easy comparison, we will provide both visual-angle and eye-angle values in this article [19].

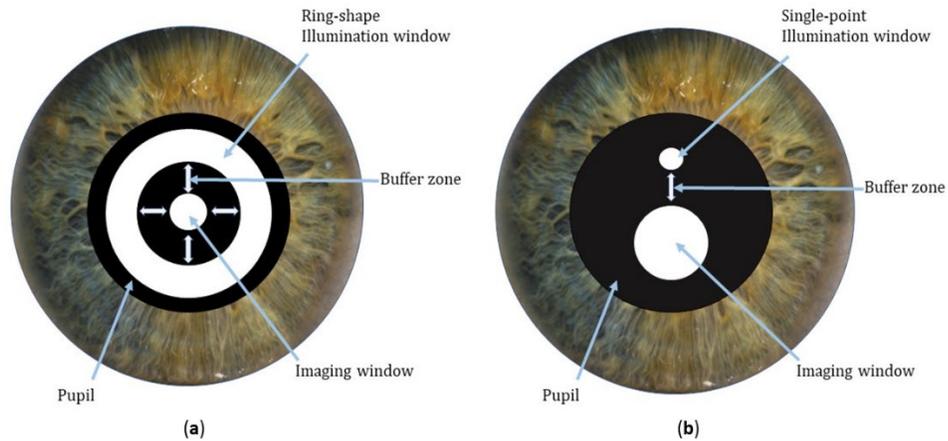

Fig.1. (a) Schematic illustration of conventional transpupillary illumination with annular pattern in traditional fundus cameras (a) and miniaturized indirect illumination with simplified single-point pattern (b).

## 2. Materials and methods

*2.1 Experimental Setup*

Figure 2 illustrates the optical layout (Fig. 2(a)) and representative photograph (Fig. 2(b)) of the system during imaging. The camera lens (CL) (8 mm focal length, F/2.5 micro video lens) and light source (LS) are situated on the same plane which is conjugated to the pupil of the eye. The ophthalmic lens (OL) (25 mm focal length) creates an image of the retina at the retina conjugate plane (RCP) which is relayed to the camera sensor (CS) (FL3-U3-120S3C-C, FLIR Integrated Imaging Solutions Inc., Richmond, Canada). The magnification from the eye pupil to the CS-LS plane is set to be 4X. Assuming the pupil diameter is 4 mm at room light condition without pupillary dilation, this corresponds to a 16 mm diameter region available in the CL-LS plane for the CS and LS to be located in. The light source contains an 810 nm LED (M850LP1, Thorlabs Inc., Newton, NJ) for near infrared (NIR) imaging guidance and a broadband (FWHM 104 nm) LED with center wavelength at 565 nm (M565D2, Thorlabs Inc., Newton, NJ) for color fundus imaging. The 565 nm illumination is optimal for visualizing retinal vasculature, differentiating the arterioles and the venules [20]. For understanding the back-reflected light distribution without polarization control, non-sequential ray tracing was implemented with Zemax OpticStudio 18.7 (ZEMAX LLC., Kirkland, WA, USA). Figure 2(c) shows the reflectance artifact pattern at camera sensor. There are two bright spots at the center with maximum intensity, as well as reflected light rays distributed throughout the whole sensor plane. Hence, apart from having two saturated spots at the center of the image, there would be an overall haze in the fundus image if the back-reflected light is not rejected. Figure 2(d) shows a fundus image taken without any polarization control where the reflectance artifacts and overall haze could be observed. Hence, orthogonal polarizers P1 and P2 (LPVISE050-A, Thorlabs Inc., Newton, NJ) are set in front of the light source and camera lens to achieve cross polarized light illumination and detection to remove back reflected light from the ophthalmic lens surface facing the camera lens.

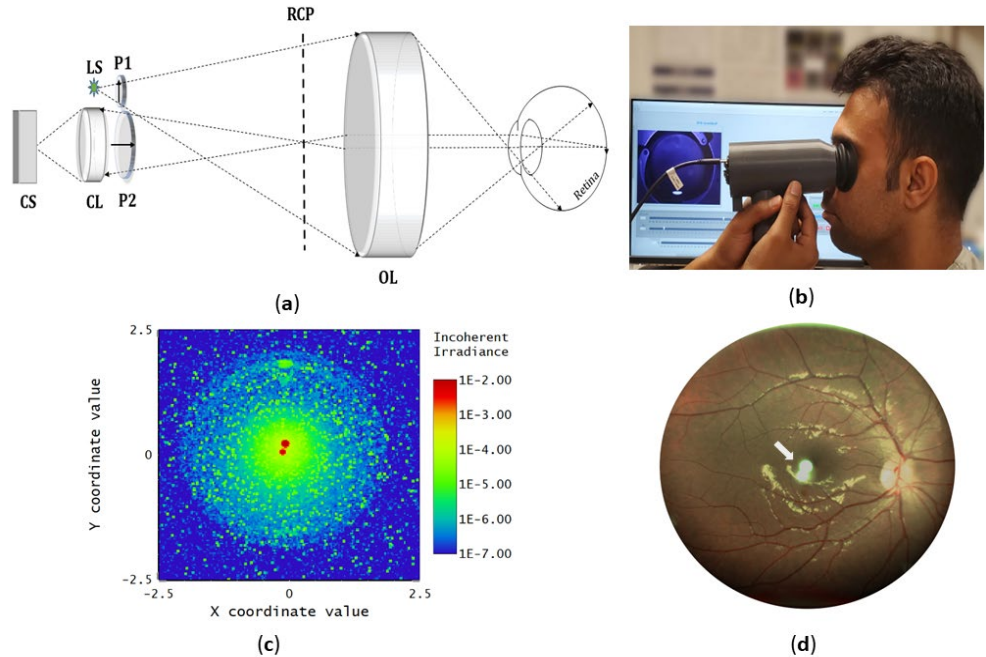

Fig. 2. (a) Optical diagram of the imaging system. (b) Photographic illustration of the imaging operation. (c) Back reflection pattern at the sensor. (d) Representative fundus image with back reflection artifact (marked with white arrow).

## 2.2 High Dynamic Range Imaging

In signal theory, the dynamic range is expressed as the ratio of the largest measurable signal to the smallest measurable signal. In the case of imaging, the signal is the light reflected from any surface and falling onto the sensor. The sensor is composed of pixels, and each pixel generates electronic signal proportional to the number of photons falling into it. If the amount of light on a pixel is equal to its full well capacity ($Q_{max}$), the pixel is saturated, and a further increase in light intensity will not change the output value. Similarly, there is a minimum amount of energy ($Q_{min}^*$) that must fall onto a pixel to generate any signal. Moreover, due to sensor's inherent noise, the actual minimum energy that causes perceptible signal output is higher, we will call it $Q_{min}$. Therefore, the dynamic range of the sensor is

$$D = \frac{Q_{max}}{Q_{min}} \quad (1)$$

It is evident that, if part of a scene reflects less energy than $Q_{min}$ or more energy than $Q_{max}$, the sensor will not be able to capture those details. The actual dynamic range is not dependent on the number of bits in the image created by the sensor but determined by the pixel size and the material property of the sensor. Since a tradeoff is inevitable among the sensor resolution, pixel size, and imaging speed, the dynamic range of a digital camera is limited. HDR imaging is a technique to extend the dynamic range of an imaging system beyond the dynamic range of the digital camera sensor.

Figure 3 illustrates basic principle of HDR imaging. The horizontal bars show the luminescence range of the scene in dim, moderate and bright illumination conditions. $LDR_1$, $LDR_2$ and $LDR_3$ are corresponding images of the scene. In dim illumination condition, the region marked by the yellow square in $LDR_1$ is below the noise floor, thus the details are imperceptible. By increasing the scene illumination, the brightness of this region could be lifted above the noise floor, shown in $LDR_3$. However, as we increase the scene illumination, region which were bright in $LDR_1$, could go above the saturation, depicted by the region marked with blue square in $LDR_3$. It is evident that, combinedly $LDR_1$, $LDR_2$ and $LDR_3$ improved the visualization of details presented in the scene, although individually they show compromised visualization due to light saturation or background noise. The HDR image combines the information preserved by the image-set and produces a composite image containing crucial details in both highlights and shadows.

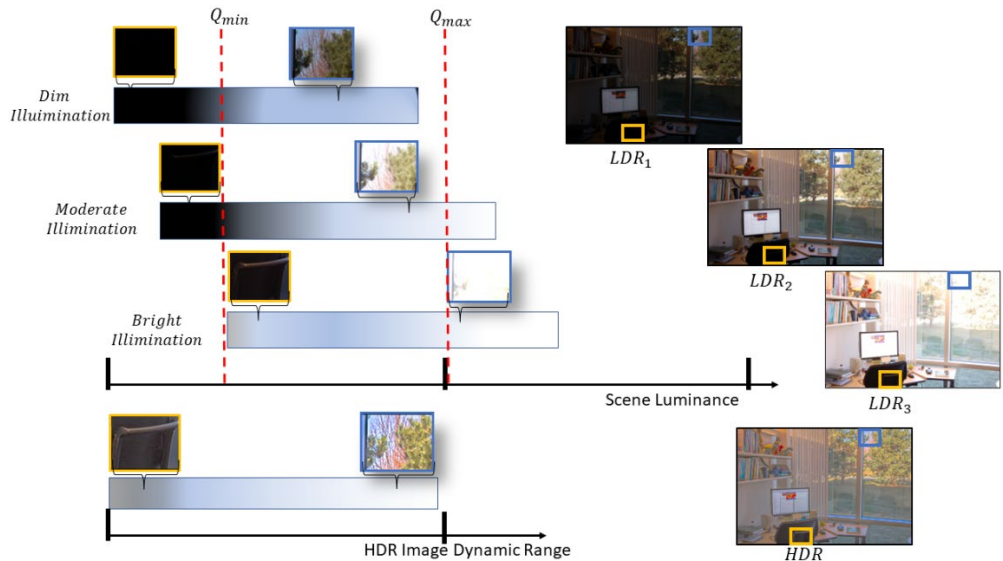

Fig. 3. Schematic illustration of the HDR imaging principle.

If N images of a scene are taken with N different illumination conditions, then the intensity value for each pixel of HDR image $L_{ij}$ can be estimated using this formula [21] :

$$L_{ij} = \sum_{k=1}^{N} \frac{f^{-1}(z_{ij})w(z_{ij})}{\Delta x} / \sum_{k=1}^{N} w(z_{ij}) \qquad (2)$$

where $z_{ij}$ is the pixel value of the LDR images and $w(z_{ij})$ is the weighting function associated with that pixel. These weights determine how much each LDR pixel would contribute to the corresponding HDR pixel. The camera response function is denoted as $f$. LDR images could be taken by varying the illumination parameters, such as exposure time, flash power, etc. $\Delta x$ denotes the parameter that is varied, and each pixel of an LDR image should be divided by the respective parameter value (exposure time or the power used to take that image) for normalization. The weighting function is calculated from the camera response function [22-24].

Theoretically, increasing the number of LDR images would reduce the effect of sensor noise and thus increase the details preserved [25]. But for nonmydriatic fundus imaging, the total acquisition time should be below the pupillary reflex time, which is around 150 ms. Therefore, we took three images of the eye, each with 35 ms exposure time, and varied the illuminating power. The rationale of changing the illumination power instead of exposure time is as follows. Firstly, there is hardware overhead delay when the exposure setting is changed between two acquisitions. Secondly, the long exposure time needed to get the high intensity LDR image causes motion blur in some cases. The power level for three images were experimentally set after calibrating with subjects from different races. Increasing the illumination power by a factor of two with each acquisition worked the best for this study. With this configuration, all the shadow, midtone and highlight information were reasonably preserved in the HDR images.

## 2.3 Human subjects and imaging protocol

This human study was approved by the Institutional Review Board of the University of Illinois at Chicago and followed the ethical standards stated in the Declaration of Helsinki. Informed consent was obtained from each subject. It was confirmed that none of the subjects had any history of seizure since the experiment involved bright flashes of light. The minimum required pupil size is 4 mm for our system, which is readily available in dimly lit room conditions. The alignment of the system and focusing was done using NIR light, so it did not stimulate the pupillary reflex. After focusing, three sequential visible light images were taken with the illumination power being doubled with each acquisition (e.g., 2 mW, 4 mW and 8 mW at the pupil plane for the subject shown in Fig. 4) with 35 ms exposure time. A LabVIEW interface was created to stream the live view of the alignment procedure and to take sequential images. After capturing the image sequence, HDR images were created. For FOV extension, a dimly lit LED fixation target was placed in front of the subjects, and they were instructed to look at the target with the eye that was not being imaged. In order to compare the quality of the images taken by our device, fundus images were taken afterwards with a commercially available portable fundus imaging device Pictor Plus (Volk, Mentor, OH).

## 2.4 Light safety

ISO standard "Ophthalmic Instruments—Fundus Cameras" (10940:2009) [26] was used to quantitatively calculate the ocular safety of the retina against photochemical hazards. The ISO standard allowed 10J/cm² radiant exposure on the retina which is 10 times lower than the retinal photochemical damage threshold. It also provides photochemical hazard weighting function to calculate the weighted irradiance. We neglect any light scattered by the cornea or the crystalline lens and assume all of the light falling at the cornea plane reaches the retina. For visible light, we calculated the weighted irradiance of each flash by considering the spectrum of the LED and the photochemical hazard function. Afterwards, we added the products of weighted irradiance and exposure time of each flash to get the radiant exposure on the retina for each acquisition. The radiant exposure of each acquisition was calculated to be $4.71 \times 10^{-6}$ J/cm², which is well below the safety limit.

For the NIR guidance illumination, weighted irradiance was calculated to be 0.06 mW/cm². We calculated the permitted time for continuous guidance using NIR wavelength with this formula:

$$T_{max} = \frac{Maximum\ allowed\ weighted\ irradiance}{Calculated\ weighted\ irradiance} \quad (3)$$

The maximum exposure time for continuous illumination with NIR light is 46.3 hours.

## 3. Results

Figure 4 illustrates the imaging procedure done on a healthy subject. The NIR guidance is illustrated in Fig. 4 (a). As the design wavelength of the polarizer is in the visible light region, the NIR image had back-reflection artifacts. Since NIR wavelength was exclusively used for guidance, this reflection artifacts were not an issue for us. Three LDR images taken with three illumination power levels are shown in Fig. 4 (b-d). The visible light images are free of any reflection artifacts.

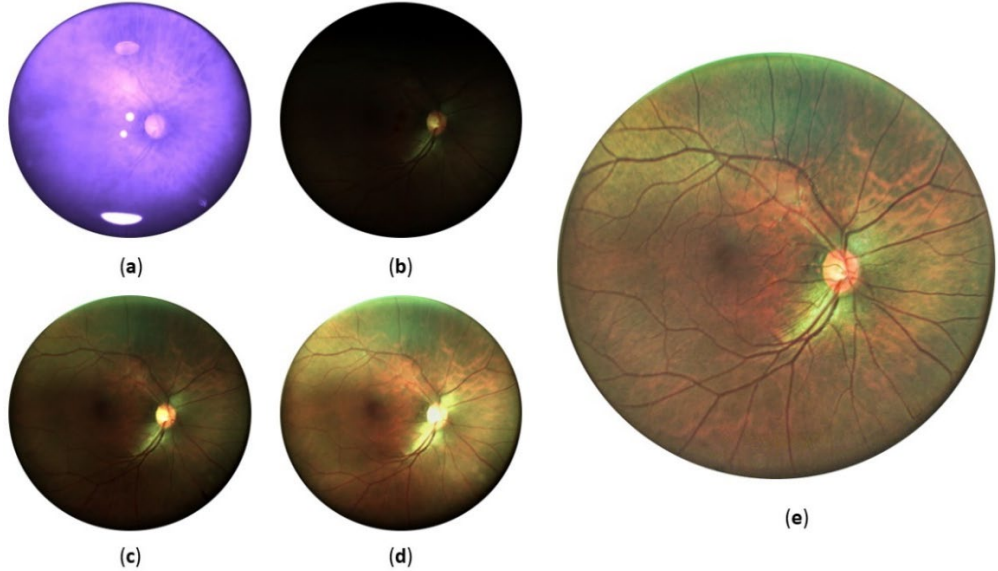

Fig. 4. (a) NIR guidance image. (b) LDR image using low power flash. (c) LDR image using medium power flash. (d) LDR image using high power flash. (e) HDR image composed of the LDR images.

Details in different fundus regions are preserved in different images of the image set. For example, the low power image shown in Fig. 4 (b) preserves the information at the optic disc, but the other regions are barely recognizable. The LDR image taken with medium power level in Fig. 4 (c) preserves the structural details of the nerve fibers around the optic disc. And finally, the high power LDR image in Fig. 4 (d) preserves the details near the macula and the periphery, but the optic disc is saturated. Notably, none of the LDR images are holistic and each excludes crucial details of the retina which are either near the noise floor or saturation. Figure 4 (e) is the HDR image created from the LDR images. All of the above-mentioned information contained by the set of LDR images are preserved in the single HDR image.

Figure 5(a) and Fig. 5(b) show the LDR images and the HDR image from a patient diagnosed with DR. A small section around the optic disc (marked with blue square) was cropped from the LDR and HDR images, shown in Fig. 5(c). Similarly, another more peripheral (marked with yellow square) region was also selected, shown in Fig. 5 (d). It is evident from Fig. 5(c) and Fig. 5(d) that the HDR image preserves the information of small vessel growth (neovascularization, marked with a black arrowhead) overlying the optic disc, as well as microaneurysms (marked with a white arrowhead) present in the periphery, giving the clinicians a comprehensive understanding of the stage of DR in this patient. None of the LDR images preserved both of this information at the same time.

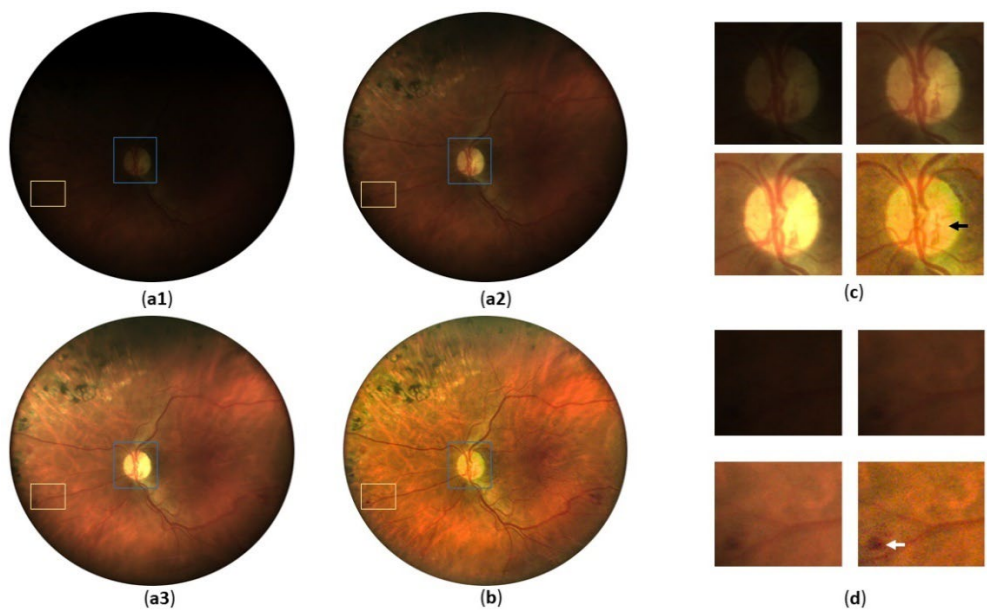

Fig. 5. (a) LDR images of an eye diagnosed with DR with different illumination power levels. (b) HDR image of the eye. (c) Cropped portion marked with blue squares in LDR and HDR images. (d) Cropped portion marked with yellow squares in LDR and HDR images.

For comparative evaluation, Fig. 6 and Fig. 7 show HDR images captured with our prototype system (Fig. 2) and a commercial portable fundus camera Volk Pictor Plus (Volk, Mentor, OH) from patients diagnosed with DR and AMD, respectively. A small portion of the fundus showing microaneurysms were cropped from Fig. 6 (a) and Fig. 6 (b) and the corresponding hot colormap is presented in Fig. 6 (c). It is evident from the colormap that the contrast of the microaneurysms (marked with black circles) is much better in the image taken with the experimental system. In a similar manner, a section around the macula of a patient diagnosed with AMD was cropped (Fig. 7). The pigment clumping (black spots) and RPE atrophy (white granular structures) are clearly more recognizable in the HDR image, evident from the respective hot colormap.

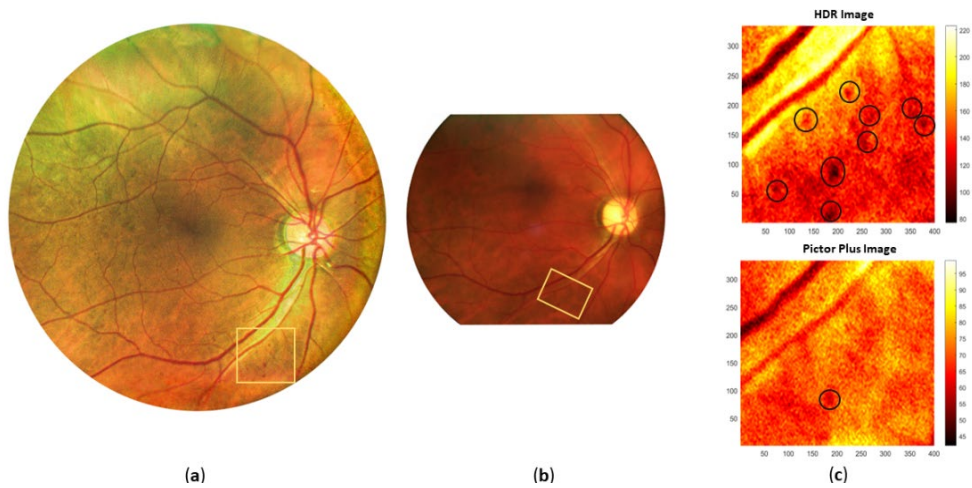

Fig. 6. (a) HDR image of a patient diagnosed with DR. (b) Image from Volk Pictor Plus from the same subject. (c) Hot colormap of the cropped regions marked with yellow squares.

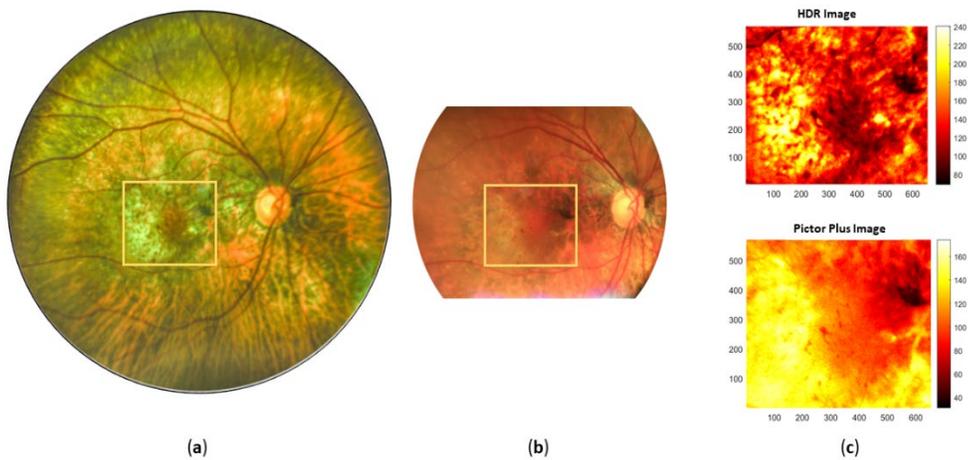

Fig. 7. (a) HDR image of a patient diagnosed with AMD. (b) Image from Volk Pictor Plus from the same subject. (c) Hot colormap of the cropped region marked with yellow square.

To demonstrate the capability of imaging peripheral retina, we took seven images with the aid of an external fixation target and merged them together (Fig. 8. (a)). A comparative image with ultra-widefield SLO Optomap is shown in Fig. 8 (b). The FOV of the image in Fig. 8(a) is estimated to be 190° eye-angle (134° -visual-angle) horizontally and 146° eye-angle (100° visual-angle) vertically. From Fig. 8 it is evident that, all major features in horizontal direction (marked with the white arrows 1-5 in both images) imaged with the SLO is also preserved in the image taken with the portable HDR fundus camera. In the vertical direction, the Optomap suffers the effect of eyelashes due to the laser scanning. Without the scanning requirement, the snapshot HDR fundus camera is not affected by the eyelashes, and thus can image the peripherical structures (yellow arrows, Fig. 8(a)) beyond the accessible region in Optomap image (Fig. 8(b)).

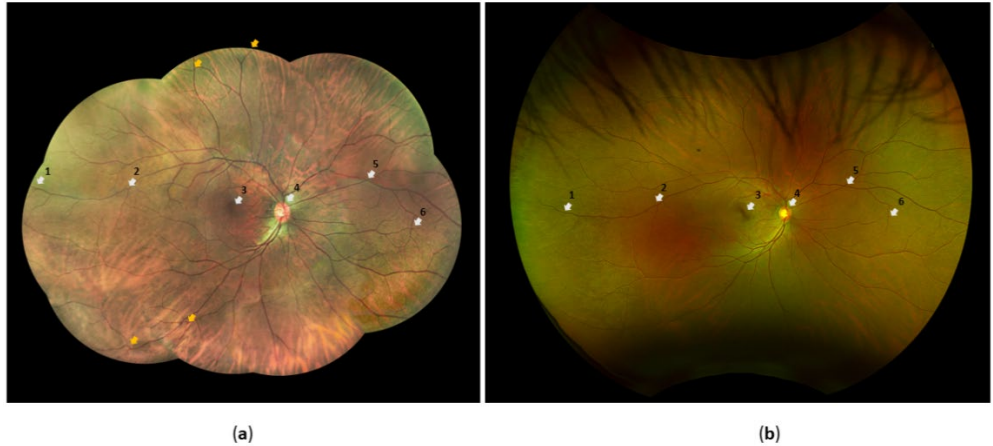

Fig. 1. (a) Ultra-widefield HDR image by merging seven HDR images together. (b) Representative image taken with Optomap.

## 4. Discussion

In summary, we have demonstrated a portable widefield fundus camera with nonmydriatic HDR imaging capability. The HDR fundus camera employed miniaturized indirect ophthalmoscopy illumination to enable the portable design for widefield fundus photography. NIR light for imaging focusing guidance and pulsed visible light illumination with flexible power control allowed rapid, nonmydriatic imaging, within a time window before visible light illumination caused pupillary response.

To overcome the limitation of FOV in traditional fundus cameras with transpupillary illumination, trans-pars-planar and trans-palpebral illumination have been demonstrated to increase the FOV up to the ora serrata, i.e., the far end of the retina [27-30]. Nevertheless, the illumination efficiency depends on the light wavelength, illumination location and the pigmentation level of the subject. Further investigation is required to standardize the trans-pars-planar and trans-palpebral and imaging protocols for clinical deployment. Miniaturized indirect illumination has been demonstrated as an alternative illumination strategy for developing widefield portable fundus cameras. However, the illumination reflectance artifact and light efficiency inhomogeneity limit the fundus image contrast. By rotating the ophthalmic lens, Toslak et. al. [17] captured two frames with lens artifacts in different locations, and merged these two frames to get an artifact free image. The moving components used in the system hindered its application as a portable device. In this study, we employed orthogonal polarization illumination and imaging control to eliminate the back reflected light. Since there was no moving component in the system, the portable fundus camera design was readily achieved with simplified indirect illumination.

For image contrast enhancement, HDR imaging has been well established to expand the dynamic range of digital imaging. For example, most of the smartphone cameras have such HDR imaging option. However, the HDR imaging has not been previously reported in fundus cameras. For regular imaging situations, the exposure time periods can be flexibly controlled to optimize the visibility of low and high brightness components in sub-frames for following HDR processing. However, for nonmydriatic fundus imaging, the available time is limited due to pupillary response caused by visible light illumination. In this study, we combined NIR light guidance and rapid visible light power control to meet the requirement for nonmydriatic fundus imaging. With the demonstrated HDR function, the portable widefield fundus camera showed superior capability to reveal pathological markers such as microaneurysms caused by DR (Fig. 6), and pigment clumping and RPE atrophy caused by AMD (Fig. 7), compared to a commercial portable fundus camera Volk Pictor Plus (Volk, Mentor, OH).

For comprehensive eye examination, fundus imaging of both central and peripheral regions is important. Our demonstrated portable HDR fundus camera has a 101° eye-angle (67° visual-angle) snapshot FOV for nonmydriatic fundus photography. In coordination with a fixation target, the FOV could be extended up to 190° eye-angle (134° visual-angle) horizontally and 146° eye-angle (100° visual-angle) vertically to visualize the retinal periphery for comprehensive eye examination. We conducted comparative imaging and analysis with the portable HDR fundus camera and an ultra-widefield SLO Optomap imager to evaluate the FOV and image contrast. As shown in Fig. 8, the portable HDR fundus camera showed similar imaging performance, and even better capability to reach peripheral fundus in the vertical direction. The widefield SLOs have been well established for improved clinical management of eye diseases, compared to traditional fundus camera. However, the high cost and bulky design make them challenging for telemedicine application, which is particularly important for rural and underserved areas with limited access to experienced ophthalmologists and expensive medical devices. A low-cost, portable, widefield fundus camera may offer a unique opportunity to foster telemedicine ophthalmology to reduce disparities of medical care in rural and underserved areas.

Although the HDR function is demonstrated with a lab prototype nonmydriatic device, we anticipate that the same HDR function can be implemented with clinical fundus cameras. In principle, multiple images are required with different light exposure controls to selectively capture the details with variable brightness for following HDR processing. For nonmydriatic fundus photography, all images should be captured before the pupillary response caused by visible light illumination. For mydriatic fundus photography, the required image-set can be readily acquired with either illumination exposure time or power control and after subsequent registering, HDR image can be generated.

## 5. Conclusion

A portable widefield fundus camera with nonmydriatic HDR imaging capability has been demonstrated and validated with both normal healthy and pathologic eyes. Miniaturized indirect ophthalmoscopy illumination was employed to achieve wide field of view, and orthogonal polarization control was used to eliminate illumination reflectance artifacts. Flash bracketed image acquisition and HDR processing were validated to implement high contrast fundus imaging. The portable HDR fundus camera provided superior capability to reveal pathological markers, compared to a conventional fundus camera. Because of the contrast improvement with HDR function, the fundus image contrast is comparable to the SLO Optomap imager. This portable, widefield, nonmydriatic HDR fundus camera promises a unique solution to facilitate affordable telemedicine.

**Funding**

National Eye Institute (P30 EY001792, R01 EY023522, R01 EY030101, R01EY029673, R01EY030842, R44 EY028786); Research to Prevent Blindness; Richard and Loan Hill Endowment.

**Disclosures**

Patent applications for wide field fundus illumination and photography.